\documentstyle[prl,aps,epsfig,floats]{revtex}
\def\DESepsf(#1 width #2){\epsfxsize=#2 \epsfbox{#1}}
\begin{document}
\pagestyle{empty}                                      
\preprint{
\hbox to \hsize{
\hbox{
            }
\hfill $
\vtop{
 \hbox{ }}$
}
}
\draft
\vfill
\twocolumn[\hsize\textwidth\columnwidth\hsize\csname @twocolumnfalse\endcsname
\title{Possibility of Large Final State Interaction Phases
in Light
of $B \to K\pi$ and $\pi\pi$ Data }
\vfill
\author{$^{1}$Wei-Shu Hou and $^2$Kwei-Chou Yang}
\address{
\rm $^1$Department of Physics, National Taiwan University,
Taipei, Taiwan 10764, R.O.C.\\
\rm $^2$Department of Physics, Chung Yuan Christian University,
 Chung-li, Taiwan 32023, R.O.C.
}

%
%
\vfill
\maketitle
\begin{abstract}
The newly observed $\overline B^0 \to \overline K^0\pi^0$ mode is
quite sizable while $\pi^-\pi^+$ is rather small.
Data also
hint at $\pi^-\pi^0 \gtrsim \pi^-\pi^+$.
Though consistent with zero,
central values of $CP$ violating asymmetries in
$K^-\pi^{+,0}$ and $\overline K^0\pi^-$ show an interesting
pattern.
Taking cue from these, we suggest that,
besides
$\gamma\equiv {\rm arg}\,(V_{ub}^*)$ being large,
the rescattering
phase $\delta$ in $K\pi$ and $\pi\pi$ modes
may be greater than $90^\circ$.
If this is true, not only the above trends can be accounted for,
one would
also find $\pi^0\pi^0 \sim \pi^-\pi^{+,0}$, and the $CP$
asymmetry
in $\overline B^0$ vs. $B^0 \to \pi^-\pi^+$ could be as large as $-60\%$.
These results can be tested in a couple of years.
\end{abstract}
\vskip2pc]

\pagestyle{plain}

The branching ratios (Br) of all four $B \to K\pi$ modes
as well as the $\pi^-\pi^+$ mode have recently been reported by
the CLEO Collaboration \cite{pipiLP}.
The measurements of $K^- \pi^+$, $K^- \pi^0$,
and $\overline K^0\pi^-$ modes have been improved,
while $\overline K^0\pi^0$ and $\pi^-\pi^+$ modes are newly observed.
Direct CP asymmetries ($a_{\rm CP}$s) in 5 charmless hadronic modes
have also been obtained \cite{acpLP} for the first time,
albeit with large errors.
The ratio $K^- \pi^+ / \overline K^0 \pi^-~\simeq~1$ suggests that
the unitarity phase angle $\gamma$ ($\equiv {\rm arg}V_{ub}^*$) in
the Cabibbo-Kobayashi-Maskawa (CKM) matrix is of order $90^\circ$.
Surveys of observed and emerging charmless rare B modes suggest that
$\gamma > 90^\circ$~\cite{DHHP,HHY,HY,HSW},
which is in some conflict with $\gamma \sim 60^\circ - 70^\circ$
obtained~\cite{parodi} from the global CKM fit
to data other than charmless hadronic B decays.
The ratio $K^-\pi^0/K^-\pi^+~\approx~0.65$ confirms
the expectation that the electroweak penguin (EWP) contribution
to the $K^- \pi^0$ mode is constructive towards
the leading strong penguin (P) contribution~\cite{DHHP,HY,CCTY}.
These results illustrate the information contained in charmless B
decays that have been emerging in the past 3 years.

However, the strength of the newly observed $\overline K^0\pi^0$ mode,
at $(14.8^{+5.9+3.5}_{-5.1-4.1}) \times 10^{-6}$,
is hard to understand, since EWP-P interference should be destructive.
One would have expected that $\overline K^0 \pi^0 / K^- \pi^0 \approx 1/3$,
or $\overline K^0\pi^0 \sim 6 \times 10^{-6}$.
The errors are still large,
but if we take the present central value seriously,
since $\overline K^0 \pi^0$ mode is only weakly dependent on $\gamma$,
a natural possibility is the presence of
strong final state interaction (FSI)
rescattering.
What is more, we find that present $a_{\rm CP}$ central values
as well as the indication that $\pi^-\pi^+ <
\pi^-\pi^0$ all offer some support for this possibility.

The long awaited $\pi^-\pi^+$ mode is finally measured at
$(4.7^{+1.8}_{-1.5}\pm 0.6) \times 10^{-6}$, which is rather small.
The data also hint at the $\pi^-\pi^0$ mode.
Though not yet significant enough,
preliminary CLEO findings give
\cite{pipiLP}
$\pi^-\pi^0=(5.4^{+2.1}_{-2.0}\pm 1.5)\times 10^{-6}$,
and the central values seem to indicate that
$\pi^-\pi^+ \lesssim \pi^-\pi^0$.
This could be brought about by large $\gamma$ and/or large FSI phase.
The two pictures can be distinguished
by measuring $\pi^0\pi^0$~\cite{HHY,CCTY}.
If small $\pi^-\pi^+$ is due to $\gamma>90^\circ$
while FSI phase $\delta\leq 30^\circ$ is small,
then $\pi^0\pi^0\lesssim 10^{-6}$ is expected.
However, if the mechanism is due to large FSI phase $\delta
>90^\circ$, then $\pi^0\pi^0$ can reach $\sim 5 \times 10^{-6}$,
i.e. as large as $\pi^-\pi^+$.

It is known that the $a_{\rm CP}$s are very sensitive to FSI phases.
Although the present accuracy of $a_{\rm CP}$s
in $K\pi$ modes is limited by statistics,
the central values may already offer us a glimpse of
the trend of the FSI phase.
As we will show, we find
a coherent picture where not only $\gamma$ is large,
but large $\delta$ is preferred as well,
{\it if} the current central values are taken at face value.
Moreover, large FSI phase $\delta$
can not only be tested by finding $\pi^0\pi^0 \sim \pi^-\pi^+$,
it can also be tested in the near future
by finding rather large $a_{CP}$ in the $\pi^-\pi^+$ mode.
The size of FSI phases in B decays is an issue of
great theoretical dispute~\cite{fsi},
which can only be settled by experiment.
We shall consider elastic $2\to 2$ rescattering phases
as the only long distance FSI phases,
returning to a more cautionary note towards the end of our discussion.
We do include short-distance rescattering phases arising from quark-gluon
diagrams. For simplicity, we shall also assume factorized amplitudes
that feed the elastic rescattering.

\begin{figure}[t!]
\centerline{
            {\epsfxsize3.4 in \epsffile{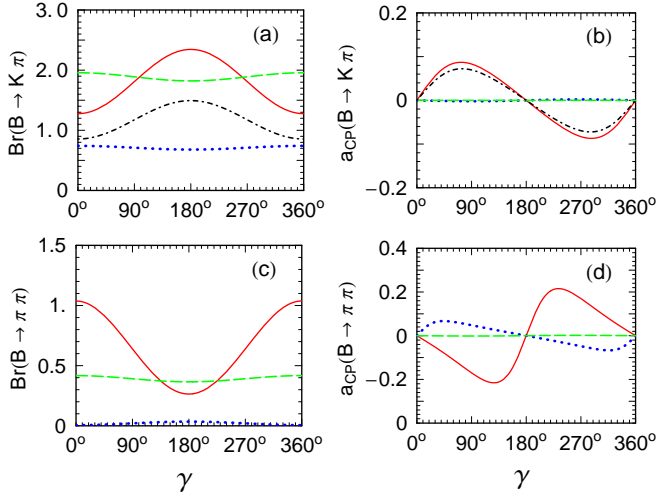}} }
\caption{
Brs and $a_{\rm CP}$s vs. $\gamma$ for $K\pi$ and
$\pi\pi$, respectively.
In (a) and (b), solid, dash,
dotdash and dots denote
$\overline B\to K^-\pi^+$, $\overline K^0\pi^-$,
$K^-\pi^0$
and $\overline K^0\pi^0$, respectively,
using $m_s = $ 80 MeV.
For
(c) and (d), solid, dash and dots represent
$\overline B\to \pi^-\pi^+$,
$\pi^-\pi^0$ and $\pi^0\pi^0$
with $m_d = 2m_u= $ 4 MeV.
We have used $\vert V_{ub}/V_{cb}\vert = 0.087$
and Brs are in units of
$10^{-5}$.
}
 \label{fig:gamma}
\end{figure}

Let us study first the $K\pi$ and $\pi\pi$ modes
without assuming long distance FSI phase.
For the relevant effective weak Hamiltonian,
we refer to Refs.~\cite{CCTY,CY,ali}.
We take $q^2=m_b^2/2$ in penguin
coefficients to generate favorably large
quark level absorptive parts~\cite{GH}.
Adopting factorization approach with $N_c=3$ and
assuming that FSI rescattering is negligible
(i.e. setting the FSI phase $\delta=0$),
the Brs and $a_{\rm CP}$s for the $K\pi$ modes
are shown vs. $\gamma$ in Figs.~\ref{fig:gamma}(a) and \ref{fig:gamma}(b).
We have rescaled the value of $F_0^{BK}=0.3$
to fit the $\overline K^0\pi^-$ data,
and take
$F_0^{B\pi}/F_0^{BK}\simeq 0.9$ for SU(3)
breaking.
We use $m_s(m_b) = 80$ MeV since lower $m_s$
improves agreement with data \cite{HHY,HSW}.
Nonfactorizable effects are usually
lumped into an effective
$N_c^{\rm eff} \neq 3$.
In our case only $\overline K^0 \pi^0$ and
$\pi^0\pi^0$ have
color suppressed tree contribution and could
be sensitive to $N_c^{\rm eff}$
if it is as small as $N_C^{\rm
eff}\sim 1$.
However,
such a small value of $N_c^{\rm eff}$ is not reasonable.

The current data give $K^-\pi^+$, $K^- \pi^0$, $\overline
K^0\pi^-$, $\overline K^0\pi^0$ $\sim$ 1.88, 1.21, 1.82, 1.48
$\times 10^{-5}$, respectively.
The observation $K^-\pi^+/ \overline K^0\pi^-\gtrsim 1$
gives a strong hint for $\gamma > 90^\circ$ if $\delta$ is negligible.
The $\overline K^0\pi^0$ rate is weakly $\gamma$-dependent
since it receives only color suppressed tree (T) contribution.
Thus, as seen from the figure, the value of $\gamma$
has little impact on the $\overline K^0\pi^0$ rate.
Without EWP, one expects both $K^-\pi^0/ K^- \pi^+$ and
$\overline K^0\pi^0 / \overline K^0\pi^-\approx (1/\sqrt{2})^2$,
where $1/\sqrt{2}$ comes from the $\pi^0$ isospin wave function.
The ratio $K^-\pi^0/ K^- \pi^+\sim 0.65$ confirms numerically
the expectation that the $K^-\pi^0$ mode is
enhanced by the EWP contribution.
However, EWP-P interference is expected to be destructive
in the $\overline K^0\pi^0$ amplitude,
hence the ratio $\overline K^0\pi^0 / \overline K^0\pi^-$
should decrease from 1/2 and read,
for $m_s=80$ MeV ($m_s$ large),
\begin{eqnarray}
\frac{\overline K^0\pi^0}{ \overline K^0\pi^-}
 \approx {1\over 2}
\left\vert 1 - r_0 \,
 {
   {3\over 2}a_9
\over   a_4 +a_6 R}\right\vert^2
 \approx  0.36\ (0.33),
\end{eqnarray}
where we have dropped the $a_2$ term for display purpose,
$r_0 = f_\pi F_0^{BK}/ f_K F_0^{B\pi} \simeq 0.9$, and
$R = 2{m_K^2/(m_b-m_d)(m_s+m_d)}$.
Small $m_s$ can enhance slightly
the $\overline K^0\pi^0/\overline K^0\pi^-$ ratio
but does not help much in understanding data.

The $a_{\rm CP}$s for $K^- \pi^{+,0}$ modes are dominated by
Im($V_{us}^* V_{ub})a_1\times $Re$(V_{ts}^* V_{tb})$\,Im($a_4+a_6 R)$
which peak at $+10\%$ around $\gamma \sim 70^\circ$.
But for $\overline K^0\pi^{-,0}$ modes,
which do not have sizable T component,
$a_{\rm CP}$s are too small to be measureable.
Due to large errors in $a_{\rm CP}$s so far,
it may be premature to compare theoretical results with data.
We note, nevertheless, that the present $a_{\rm CP}$ data~\cite{acpLP}
give the central values for $K^-\pi^+$, $K^- \pi^0$ and
$\overline K^0 \pi^-$ as $\sim$ $-0.04$, $-0.27$ and
$+0.17$, respectively,
which are not consistent with theoretical expectations of Fig. 1(b).

The T-P interference for $\pi^- \pi^{+}$
is anticorrelated \cite{HHY,SilvaWolf}
with the
$K^-\pi^{+,0}$ case
because the penguin KM factors
are
${\rm Re}(V_{td}^* V_{tb})\cong
A\lambda^3(1-\rho)$
and ${\rm Re}(V_{ts}^* V_{tb})\cong
-A\lambda^2$,
which are opposite in sign since $1-\rho > 0$ by definition.
Thus, $K^-\pi^{+,0}$ are enhanced for $\cos \gamma <0$
while $\pi^- \pi^{+}$ is
suppressed,
as illustrated in Figs. 1(a) vs. 1(c).
Although experimental error for $\pi^-\pi^0$
mode is rather large,
the central value \cite{pipiLP} seems to
suggest $\pi^- \pi^+ \lesssim \pi^- \pi^0$,
hence $\gamma
\gtrsim 140^\circ$ is favored if FSI can be neglected.
The
$a_{\rm CP}$ for $\pi^-\pi^+$ is given in Fig. 1(d),
which clearly is opposite in sign to $K^-\pi^{+,0}$.
It could be as large as $-20\%$ at $\gamma\sim 120^\circ$.
For this $\gamma$ value, one would expect
$K^- \pi^+ : \overline K^0\pi^- :
 K^- \pi^0
:\overline K^0\pi^0
\simeq 1:0.86:0.63:0.33$,
which however
deviates considerably from the present observation
of $\sim
1:0.97:0.64:0.79$, mainly in $\overline K^0\pi^0$.

\begin{figure}[t!]
\centerline{ 
            {\epsfxsize3.4 in \epsffile{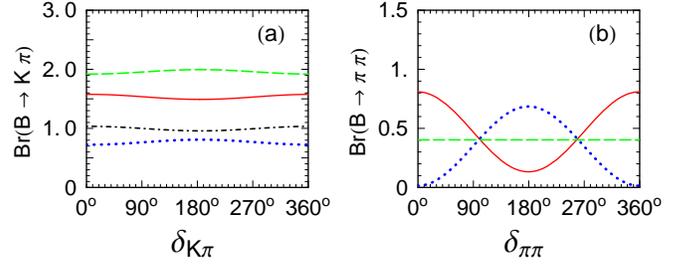}}}
\smallskip
\caption {Brs for $K\pi$ and $\pi\pi$ vs. $\delta$ for
$\gamma=64^\circ$
with notation as in Fig. 1.}
 \label{fig:gamma64}
\end{figure}

Since large $\gamma$ is favored but cannot explain
data completely,
could large
FSI phase $\delta$ {\it alone} work?
In Fig.~\ref{fig:gamma64}, we show the Brs of $K\pi$ and
$\pi\pi$ vs. $\delta$ with $\gamma=64^\circ$ \cite{parodi}.
Large $\delta \gtrsim 100^\circ$ could help explain \cite{HHY,CCTY}
$\pi^-\pi^+ \lesssim \pi^-\pi^0$, but it fails badly in the
$K\pi$ modes
since $K^- \pi^+ / \overline K^0\pi^- <0.8$,
while $\overline K^0 \pi^0$ is only half of $K^-\pi^+$
and actually gains
little from the $K^-\pi^+$ mode
via FSI rescattering.
We therefore conclude that large $\delta$ alone cannot account for data
and large $\gamma$
is still favored,
but large $\delta$ in this context should be further explored.

Before studying the case of having $\gamma$ and $\delta$ both large,
let us make explicit our treatment of FSI phases
in $\pi\pi$ and $K\pi$ final states.
Following the notation of~\cite{GHLR},
we decompose the $B\to \pi\pi$
amplitudes as,
\begin{eqnarray}
 {\cal A}(\overline B^0\to \pi^-\pi^+)
 &=&A_0e^{i\delta_0}+A_2 e^{i\delta_2},\nonumber\\
 \sqrt{2}{\cal A}(\overline B^0\to \pi^0\pi^0)
 &=&A_0e^{i\delta_0}-2A_2 e^{i\delta_2},\nonumber\\
 \sqrt{2}{\cal A}( B^-\to \pi^-\pi^0)
 &=&3A_2 e^{i\delta_2},
 \end{eqnarray}
where $A_{0,2}$ correspond to final state isospin 0 and 2,
and $\delta_{0,2}$ are FSI phases.
For $K\pi$ modes, we decompose the
amplitudes into
$A_{3/2 (1/2)}$ for
$\Delta I =1$ transitions to
$I=3/2 (1/2)$ final states,
and $B_{1/2}$ for
$\Delta I =0$ transitions to $I=1/2$ final states,
\begin{eqnarray}
 {\cal A}(\overline B^0\to K^-\pi^+)
 &=&A_{3\over 2} e^{i\delta_{3\over 2}}
  -(A_{1\over 2}-B_{1\over 2})  e^{i\delta_{1\over 2}},\nonumber\\
 \sqrt{2}{\cal A}(\overline B^0\to \overline K^0\pi^0)
 &=&2A_{3\over 2} e^{i\delta_{3\over 2}}+(A_{1\over 2}-B_{1\over 2}) e^{i\delta_{1\over 2}},\nonumber\\
 {\cal A}( B^-\to \overline K^0\pi^-)
 &=&-A_{3\over 2} e^{i\delta_{3\over 2}}+(A_{1\over 2}+B_{1\over 2})  e^{i\delta_{1\over 2}},\nonumber\\
  \sqrt{2}{\cal A}( B^-\to K^-\pi^0)
  &=&2A_{3\over 2} e^{i\delta_{3\over 2}}+(A_{1\over 2}+B_{1\over 2})  e^{i\delta_{1\over 2}},
\end{eqnarray}
and $\delta_{{3\over 2},{1\over 2}}$ are FSI phases.
Short distance quark-antiquark rescattering effects have been included,
which lead to small and calculable perturbative phases
for $A_{3\over 2}$, $A_{1\over 2}$ and $B_{1\over 2}$.
Because in $K\pi$ modes the strength of EWP
is comparable to T,
it is known \cite{DH} that SU(3)
relations of Ref. \cite{GHLR} do not hold.
We refrain from discussing SU(3) but restrict ourselves to
SU(2) (isospin), where $\delta_{i}$ in
Eqs. (2) and (3) are nominally elastic FSI phases
but they model long-distance rescattering.
The phase differences are observable, which we denote as
$\delta_{K\pi}=\delta_{3\over 2}-\delta_{1\over 2}$ and
$\delta_{\pi\pi}=\delta_2-\delta_0$
.
Unlike the aforementioned case of EWP $\sim$ T
in the amplitudes $A_i$ and $B_i$,
electroweak effects in FSI rescattering are negligible
compared to strong FSI.

\begin{figure}[t!]
\centerline{
            {\epsfxsize3.4 in \epsffile{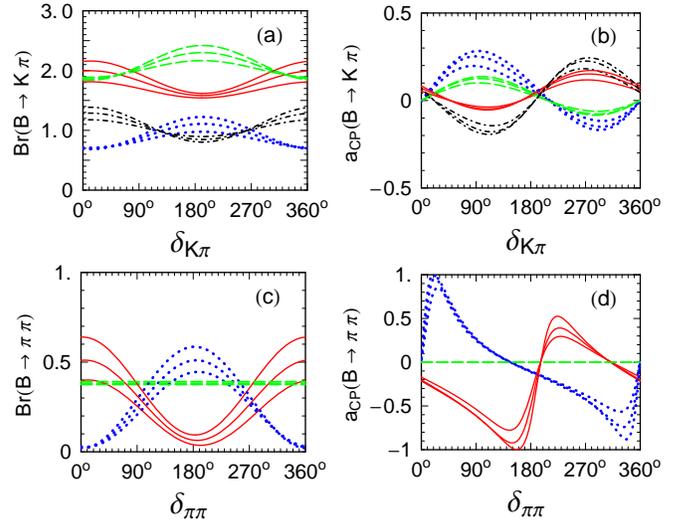}}}
\smallskip
\caption {Brs and $a_{\rm CP}$s for $K\pi$ and $\pi\pi$ vs.
$\delta$.
For curves from:
 (a) up (down) to down (up) for
$K^-\pi^+, \overline K^0\pi^{-,0}$ ($K^-\pi^0$) at
$\delta=180^\circ$; and
 (b) up (down) to down
(up) for $K^-\pi^{+,0}$
    ($\overline K^0\pi^{-,0}$) at
$\delta=90^\circ$; and
 (c) down to up at $\delta=180^\circ$; and
 (d) down to up for $\pi^-\pi^+$ ($\pi^0\pi^0$)
     at $\delta=160^\circ$ ($20^\circ$),
are for
$\gamma=130^\circ, 110^\circ$, and $90^\circ$, respectively.}
\label{fig:gammalarge}
\end{figure}

We plot in Fig.~\ref{fig:gammalarge} the Brs and
$a_{\rm CP}$s vs. $\delta$
for $K\pi$ and $\pi\pi$, respectively, for several large $\gamma$ values.
From Fig.~3(a), we see that $K^-\pi^+ \to
\overline K^0 \pi^0$
FSI rescattering is magnified by large $\gamma$,
while
$K^-\pi^0\to \overline K^0 \pi^-$ rescattering
enhances $\overline
K^0 \pi^-$.
This is because for $\cos\gamma < 0$,
the T
contribution (hence $A_{3\over 2}$) changes sign,
leading to a
marked change in the rescattering pattern.
Taking $\gamma=110^\circ$ and $\delta_{K\pi}=90^\circ$,
we obtain $K^- \pi^+ :
\overline K^0\pi^- : K^- \pi^0
                     : \overline K^0\pi^0 \simeq
1:1.1:0.61:0.47$,
and the relative size of $\overline K^0\pi^0$ has
been enhanced by $\sim 30\%$.
Such enhancement occur only when
$\delta_{K\pi} \gtrsim 60^\circ$.
Note that with larger $\gamma$, say
$130^\circ$,
$K^- \pi^+ \simeq \overline K^0\pi^-$ is still
possible if $\delta_{K\pi} \simeq 90^\circ$.
For even larger $\delta_{K\pi}$, in principle one can have
$\overline K^0\pi^0 > K^- \pi^0$
but then $\overline K^0\pi^- > K^- \pi^+$ as well.
The preferred combination of $\gamma$ and $\delta_{K\pi}$ can be
better determined as data improve.
For the $a_{\rm CP}$s, as
shown in Fig.~\ref{fig:gammalarge}(b), one has
$K^- \pi^+ : \overline K^0\pi^- : K^- \pi^0
:\overline K^0\pi^0
\sim$ $-0.04\ (-0.04),\
0.13\ (0.17)\ , \ -0.16\ (-0.27),\ 0.2$,
respectively, for $\gamma=110^\circ$ and $\delta_{K\pi}=90^\circ$,
where the numbers in
parentheses are
the experimental central values.
Although
these numbers should not be taken too seriously at present,
we note that if current experimental
$a_{\rm CP}$ central values continue to hold,
they can be accounted for by having $\gamma$ and $\delta_{K\pi}$ both large.
Without final state rescattering,
the $a_{\rm CP}$s for
$K^- \pi^{+,0}$ are positive
and very close to each other, while
the $a_{\rm CP}$s for
$\overline K^0 \pi^{-,0}$ would be practically zero.
As illustrated here, with large
final state rescattering,
the $a_{\rm CP}$s for $K^- \pi^{+,0}$
change sign,
while $\overline K^0 \pi^{-,0}$ modes gain $a_{\rm CP}$s
that are opposite to
$K^- \pi^{+,0}$ modes.
These trends can be tested in the near future.

For the $\pi\pi$ modes, as shown in Fig.~\ref{fig:gammalarge}(c),
$\delta_{\pi\pi}\gtrsim 90^\circ$ could give
$\pi^-\pi^+ \lesssim \pi^-\pi^0$, which is hinted by present data.
For larger $\gamma$ values, one has
less need for large $\delta_{\pi\pi}$ phase.
Because FSI $\pi^-\pi^+ \to \pi^0 \pi^0$ rescattering
feeds the $\pi^0\pi^0$ mode,
$\pi^0 \pi^0 \sim \pi^-\pi^+ \lesssim
\pi^-\pi^0$ becomes possible.
Observing large $\pi^0\pi^0$ would be a
good indication \cite{HHY,CCTY} for large FSI.
The $a_{\rm CP}$s offer an even better test.
We plot in Fig.~\ref{fig:gammalarge}(d)
the $a_{\rm CP}$s for $\pi\pi$ modes.
For $\delta_{\pi\pi}\sim 90^\circ$,
the $a_{\rm CP}$ in $\pi^-\pi^+$ can reach $\sim -60\%$,
which is 2--3 times larger than
the case without the $\delta_{\pi\pi}$ phase.
We stress once again that, from Fig.~2, although a large
$\delta_{\pi\pi}\gtrsim 90^\circ$ with small $\gamma\sim 64^\circ$ may
explain the small observed $\pi^-\pi^+$ rate and $\pi^-\pi^+
\lesssim \pi^-\pi^0$,
the ratio $K^- \pi^+ /\overline K^0\pi^- \sim 3/4$ is
not very sensitive to $\delta_{K\pi}$
and would be too low compared to what is observed.

Before we conclude, we comment on some
uncertainties in the present study.
First, in factorization approach,
$m_s$ always appears together with $a_6$.
We have used the set of
effective Wilson coefficients of \cite{CY}.
To fit, for example the $K^-\pi^0$ mode to data,
a larger $|a_6|$ would be accompanied by
a larger $m_s$ and vice versa.
The set of $a_i$s adopted does not
change our conclusions.
Second, the $\pi^-\pi^0$ rate is insensitive
to $\gamma$
and independent of $\delta$,
and is proportional to
$\vert F^{B\pi}_0 V_{ub}\vert^2$.
Although our $\pi^-\pi^0$
result is below the present
experimental central value, the latter is not yet firm and
we have just concerned
ourselves with the $\pi^-\pi^0/\pi^-\pi^+$ ratio.
Third, factorization in the $\pi\pi$ modes has been shown \cite{BBNS}
to follow from pQCD in the heavy quark limit.
One could extract the
effective $N_C$ from this study~\cite{CY}.
However, there are no significant changes
in $K\pi$ and $\pi\pi$ modes without considering
the long-distance FSI phase $\delta$,
which is the position taken in Ref. \cite{BBNS}.
Note that the strong phases generated by
hard gluon rescattering off the spectator quark calculated in \cite{BBNS}
is destructive with the hard quark-antiquark rescatterings
in the penguin loop, resulting in
$a_{\rm CP}(\pi^-\pi^+) \sim -4\% \times \sin \gamma$,
which is much smaller than shown in Fig. 1(d).
Thus, $\pi^0\pi^0\sim \pi^-\pi^{+,0}$ and
$a_{\rm CP}(\pi^-\pi^+)$ as large as $-60\%$ would definitely indicate
the existence of large (elastic) FSI phase $\delta$
arising from {\it long distance} effects,
something that is argued \cite{BBNS} to be ($1/m_B$) power suppressed.
That would be rather interesting,
since Regge \cite{fsi} and other \cite{SW} arguments
give phase differences of order $10^\circ$-$20^\circ$.
As B factories at SLAC and KEK have already turned on,
together with the recent commissioning of the CLEO III detector at Cornell,
one should have at least ten times the present data in $2$ years.
It is
thus very likely that the FSI effects discussed
here would be tested in the near future.
Finally, we should caution that our illustration with
elastic strong FSI phase difference $\delta$ is
only phenomenological.
While $\delta \sim 90^\circ$ is in principle possible \cite{fsi},
FSI in B decays are expected to be highly inelastic \cite{SW}.
A large strong phase could well be accompanied by
deviations in the {\it magnitude} of amplitudes
from factorized ones that we have used.

In conclusion, large $\gamma$ is favored by data if
one considers $K\pi$ together with $\pi\pi$ data.
Although a large FSI phase $\delta\gtrsim 90^\circ$
with the current $\gamma\sim 64^\circ$
can account for the smallness of $\pi^-\pi^+$,
it fails to explain the observed $K^-\pi^+ / \overline K^0 \pi^-\sim 1$.
The strength of the observed $\overline K^0 \pi^0$ mode,
the hint that $\pi^- \pi^0 \gtrsim \pi^-\pi^+$,
together with the current experimental central values for
$a_{\rm CP}$ in the $K\pi$ modes,
all seem to suggest that on top of $\gamma \gtrsim 100^\circ$,
the long-distance phase $\delta$ could be as large as $90^\circ$
for both $K\pi$ and $\pi\pi$ modes.
If true, it would not only uphold the above indication and hints,
one would find an enhanced $\pi^0\pi^0$ rate comparable to $\pi^-\pi^+$,
and $a_{\rm CP}$ in the latter mode as large as $ - 60\%$.
These results should be testable in the next 2 years.

\vskip 0.5cm
\noindent{\bf Acknowledgement}.\ \
This work is supported in part by
the National Science Council of R.O.C.
under Grants NSC-89-2112-M-002-036
and NSC-89-2112-M-033-010.
We thank M.~Suzuki for useful communications.

\end{document}